\documentclass[twocolumn,showpacs,preprintnumbers,amsmath,amssymb,pra]{revtex4}
\usepackage{txfonts}
\usepackage{graphicx}
\usepackage{dcolumn}
\usepackage{bm}
\begin{document}


\title{Coherent  control of dissipative dynamics in a periodically driven lattice array}
\author{Zhao-Yun Zeng$^{1,2}$}
\author{Lei Li$^{2}$}
\author{Baiyun Yang$^{2}$}
\author{Jinpeng Xiao$^{2}$}
\author{Xiaobing Luo$^{1,2}$}
\altaffiliation{Author to whom any correspondence should be addressed: xiaobingluo2013@aliyun.com}
\affiliation{$^{1}$Department of Physics, Zhejiang Sci-Tech University, Hangzhou, 310018, P. R. China}
\affiliation{$^{2}$School of Mathematics and Physics, Jinggangshan University, Ji¡¯an 343009, P. R. China}

\date{\today}

\begin{abstract}
We find a different mechanism for suppression of decay in an open one-dimensional lattice system,
which originates from a dark Floquet state, a sink state to which the
system is asymptotically driven, whose overall probability is determined only by the parameters of the periodic driving field. The zero-quasienergy of  dark Floquet state has been shown to be not a real zero, but a vanishingly small negative imaginary number which will cause undesirable physical effect in long-time evolution of quantum states, which is extremely different from the conservative counterpart.
Another important finding is that the value of the system's effective decay, determined by the size of the non-zero imaginary part of the dark-Floquet-state-related quasienergy,
 depends not on how many localized lossy sites there are but on which one of the lossy sites is nearest to the driven site. Thus, for specially designed local dissipation, by controlling the driving parameters, it is possible for us to drive the system to a dark Floquet state with a much lower level of
overall probability loss as compared to the undriven case and with good stability over enough longer evolution time. These results are applicable to the multisite lattice system with an odd number of sites and may be significant for long-time control of  decay in a vast family of multistate physical systems with localized dissipation.

\end{abstract}
\maketitle
\section{introduction}
Understanding the influence of dissipation on the
dynamics of physical systems is both of fundamental and technological importance and has triggered a plethora of exciting investigations. In general, dissipation is considered as an undesirable destructing factor for a long-time coherent control of quantum
states. However, in recent decades it has been recognized that dissipation can play constructive roles
in tuning properties of the system and therefore enables an additional way of steering the dynamics
of quantum systems\cite{Witthaut1,Kraus,Krauter,Barmettler}. In view of the striking experimental advances on the single-site addressability in
the optical lattices where the loss can be made truly localized in selected sites\cite{Gericke,Bakr}, ultracold atoms in optical lattices with localized dissipation provide a distinguished
model system for the study of fully
governable open quantum systems\cite{Barontini}. These systems with the dissipation process have been investigated
from a nonlinear dynamics viewpoint based on a mean-field
approximation\cite{Livi,Pardo,Ng,Brazhnyi}, where the loss is introduced  as negative
imaginary chemical potential, giving origin to stable
dissipative structures and diverse nonlinear excitations such as dynamical breathers and dissipative solitons. At the same time, the dissipative dynamics of interacting system have also been  studied in terms of the master equations beyond mean-field treatment\cite{Shchesnovich1,Shchesnovich2,Witthaut2,Tomadin,Diehl,Witthaut3,Hofstetter}.
So far, it has been shown both theoretically and experimentally that the suppression of atom
losses can be achieved under the increase
of the spatially localized dissipation (interpreted also as  the quantum Zeno
effect)\cite{Barontini,Shchesnovich1,Witthaut2, Hofstetter}, and that the system stability can be improved by a suitable amount of dissipation\cite{Witthaut2, Witthaut3}.

Very recently, Guo \emph{et al.} have considered the dissipative dynamics of a single particle confined to  a three-site system with local loss from the central site, and found that the system will asymptotically evolve into one of its eigenstates
(i.e., dark state) and meanwhile its total probability will also gradually decrease to half of the initial value which then remains unchanged\cite{Guo}.
Li \emph{et al.} have studied the localization in a periodically
driven nonlinear three-site system with loss acting on the end site, and reported two different types of localization: chaos-related localization and loss-induced localization\cite{Li}. The two  above-mentioned research papers both dealt with purely dissipative systems without additional gain mechanisms. However, there is difference between them: in the former case, no matter how long the evolution time is, the very existence of dark state will prevent  the full leakage of wave packet, while in the latter case, the inhibition of decay induced by localized dissipation only exists in a certain evolution time rather than an infinite time. In addition, it has also been
reported that increasing the loss will lead to a dramatically increased localization in some specially
designed  parity-time symmetric systems\cite{AGuo}.

On the other hand, control of quantum states via periodically
oscillating external field has been one of the subjects of
long-lasting interest due to its potential application in quantum-based technologies\cite{Hanggi}.
The coherent destruction of tunneling (CDT) is one of the seminal effects
in this field, upon the occurrence of which the
driven site would be decoupled from the undriven ones in the lattice chains provided that the
system parameters are carefully chosen\cite{Grossmann}. One may naturally expect that applying  CDT effect to
dissipative quantum systems will give rise to decoupling between the lossy and lossless sites, thereby
protecting the particle at lossless site from dissipation. However, because
 CDT occurs only at isolated system
parameters, it is impossible to precisely define
a parameter point and hence genuine CDT through experiment.
Thus, one significant disadvantage of the application is that the pseudo-CDT can not really prevent the particles from reaching the leaking sites.
 Recently, a
 quantum phenomenon called dark
Floquet state, with zero quasienergy and
negligible population at all the even-number-sites, has been predicted in periodically driven three-site (or odd-$N$-site) systems\cite{Luo1}, which can serve as an alternative tool for coherent quantum control\cite{Luo1,Luo2,Luo3}.
 Generally, from previous studies\cite{Luo1,Luo2,Luo3}, it can be concluded that the population of dark Floquet state at the even-number-sites is actually not zero but an extremely small number, and especially, only in the high-frequency limit (the case that the driving frequency goes to infinity) can the corresponding quasienergy perfectly equal to zero. As to conservative systems, it is known that such an exceedingly small population at the even-number-sites can not build up physical effect on the long-time evolution of quantum states.  Yet,  what effects  dark
Floquet state will have in  the dissipative quantum systems is still an open issue which deserves to be addressed.

In this paper, we have studied how the controlled  localized dissipation and periodic driving  influence the decay dynamics of  a
single quantum particle initially prepared in the left-end
site of
one-dimensional quantum lattice systems.
We begin the discussion with the simplest three-site system in which the loss is localized on the central site and the periodic driving is applied only to the first site. In the high-frequency region, we have derived the approximate analytical solutions
of the Schr\"{o}dinger equation of the three-site system, which indicates that the system will be eventually driven into a dark Floquet state whose overall probability is determined only by the driving parameters, not by the strength of loss coefficient.
By controlling the driving parameters, it is possible for us not only to tune the transition from  oscillating decay to overdamping decay in precise manner,  but also to produce a much lower level of
overall probability loss as compared to the undriven case.
A more careful numerical analysis shows that the quasienergy of associated dark Floquet state is not a real zero but an extremely small negative imaginary number, which is intrinsically connected with a small nonzero population at the lossy central site, suggesting that the small leakage will persist indefinitely. The size of the non-zero imaginary part of the quasienergy determines the
temporal length of the system staying in the corresponding dark Floquet state, and also reflects the robustness of the proposed scheme concerning  long-time control of decay in open quantum systems. This is a quite distinct
situation from the conservative three-site system, where the population of  dark Floquet state (with a nearly zero quasienergy) at the central site is not zero, but a extremely small value which will instead cause no physical effect in long-time evolution of quantum states. These results offer informative
starting points for considering the effects of dark Floquet state in dissipative lattice systems, and similar behavior and conclusions
can be obtained for the multisite systems with other odd numbers of sites, where the loss is localized on the even $n$th site and the periodic driving is applied only to the left-end site.
In these multisite  systems, the effect of
dark Floquet state manifests itself as it does
in the three-site system: the system will be eventually driven into a dark Floquet state whose overall probability is determined  by the driving parameters.

Another remarkable finding is that the system's effective decay depends not on how many lossy sites there are but on which of lossy sites is the first lossy lattice site from the driven left-end site.
When the first lossy site from the left is far enough away from the driven left-end site, the magnitude of the imaginary part of quasienergy corresponding to dark Floquet state  will become vanishingly small, and the period of time for the system staying in the dark Floquet state without decay will become exceedingly long. Thus, the interplay between the periodic driving and localized dissipation allows the possibility to drive the system to a dark Floquet state with much higher level of total probability (lower probability loss) and with good stability over enough longer evolution time. On the other hand,
we find that the cooperated effects of CDT and dark state can be used to effectively suppress the probability decay in the long-time dynamics of the multisite lattice system with an even number of sites.

 \section{Model system}
\begin{figure}[htb]
\includegraphics[width=0.48\textwidth]{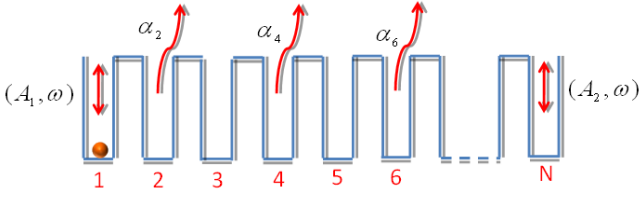}
\caption{(Color online) Schematic representation of the setup. Here the left-end site is driven periodically with amplitude
$A_1$ and frequency $\omega$, and the right-end site driven with amplitude
$A_2$ but with the same frequency. The system is subjected to localized losses from the even $n$th ($n\neq N$, that is, the right-end site is lossless irrespective of the value of $N$) sites with loss (dissipation) coefficient $\alpha_n$.}
\label{fig1}
\end{figure}

 We consider here  the dissipative dynamics of a single
quantum particle hopping on a lattice chain comprising $N$ sites,
with the end sites driven by an external periodic
field, as shown in Fig.~\ref{fig1}. In our setup, only the even $n$th sites ($n\neq N$, i.e., excluding the right-end site when $N$ is even) are designed to be subjected to losses
with dissipation strength $\alpha_{n}$ (here the index $n$ is even number). Meanwhile, we assume that the two end sites are driven
with  amplitudes $A_1$ and $A_2$ respectively, but with the same frequency $\omega$. The high tunability of external field makes it possible for us to adjust the driving parameter, and by doing so it can readily make one of the two end sites driven and the other undriven. In the tight-binding
approximation and assuming a coherent dynamics, the single-particle
motion can be generally described by the tight-binding Hamiltonian

\begin{align}  \label{equ:H}
H=&\sum\limits_{n = 1}^{N }\epsilon_n(t)\left| n \right\rangle \left\langle n \right| -v \sum\limits_{n = 1}^{N - 1} (|n\rangle\langle n+1|+|n+1\rangle\langle n|)\nonumber\\
&-i \sum\limits_{n= 1}^{N - 1}\left\{[1+(-1)^n]\frac{\alpha_n}{2}\right\}|n\rangle\langle n|,\nonumber\\
 \epsilon_1 (t)=&A_1\cos(\omega t),~~\epsilon_N (t)=A_2\cos(\omega t),~~\epsilon_n (t)=0 (n\neq 1,N),
 \end{align}
where $|n\rangle$ represents the Wannier state localized in the $n$th site,
$v$ is the coupling strength connecting nearest-neighboring
sites, and the negative imaginary part of site energies denotes the effective loss with loss coefficient $\alpha_{n}$ acting only on the even $n$th $(n\neq N)$ sites.

In the Wannier basis
representation, the quantum state of system (\ref{equ:H}) can be expanded in the form $|\psi (t)\rangle =
\sum_{n=1}^{N} {c_n (t)\left| n \right\rangle }$, where $c_n(t)$ represents
complex amplitude for occupation of the $|n\rangle$ Wannier state. From the Schr\"{o}dinger equation $i\partial_t|\psi (t)\rangle=H|\psi (t)\rangle$, the
evolution equation for the probability amplitudes $c_n(t)$ reads
\begin{eqnarray}\label{c_n}
i\frac{dc_n}{dt}&=&\epsilon_1(t)c_1+\epsilon_N(t)c_N-v(c_{n-1}+c_{n+1})-i[1+(-1)^n]\frac{\alpha_n}{2}c_n,\label{coupledN}\nonumber\\
\end{eqnarray}
where $c_{n\leq 0}=c_{n>N}=0$ and $\alpha_N=0$.

In this work, we mainly illustrate how the decay of a
single particle initially localized at the left-end site can be
suppressed  by the interplay of periodic driving and controllable localized dissipation,
with focus on the following two cases: (i) when the total site number $N$ is odd, the periodic driving is acting only on the left end; (ii) when the total site number $N$ is even, the periodic driving is acting only on the right end.

\section{control of dynamics by periodic driving and localized dissipation}
\subsection{three-site system}
First, we  take into account the control of dynamics by  driving and dissipation in the three-site model, which is the simplest quantum system  for our consideration. In this case,
the dynamical equations are of the form
\begin{align}\label{Model for c}
 i\frac{\partial}{\partial t}\left(
                                     \begin{array}{c}
                                       c_1 \\
                                       c_2 \\
                                       c_3 \\
                                     \end{array}
                                   \right)
 =\left(
    \begin{array}{ccc}
      A_1\cos(\omega t) & -v & 0 \\
      -v & -i\alpha_2 & -v \\
      0 & -v & A_2\cos(\omega t) \\
    \end{array}
  \right)\left(
           \begin{array}{c}
             c_1 \\
             c_2 \\
             c_3 \\
           \end{array}
         \right).
\end{align}

It is difficult for us to obtain exactly analytical solutions
of Eq.~\eqref{Model for c} except for the undriven case.
However, in the high-frequency regime $\omega\gg v$, the Schr\"{o}dinger equation \eqref{Model for c} can be
investigated analytically by using the high-frequency approximation method. To that end,
we introduce the transformation $c_1=a_1e^{-i\int A_1\cos(\omega t)dt}$, $c_2=a_2$,$c_3=a_3e^{-i\int A_2\cos(\omega t)dt}$,
where $a_i(t)$ are slowly varying functions.
Using  the Fourier  expansion $\exp(\pm ik\sin\omega t)=\sum_n J_n(k)\exp(\pm int)$ in term of
$n-$order Bessel functions $J_n(k)$, and neglecting all of the
orders except $n=0$ in the
high-frequency limit, we arrive at the effective equations of
motion,
\begin{align}\label{Model for a1}
 i\frac{\partial}{\partial t}\left(
                                     \begin{array}{c}
                                       a_1 \\
                                       a_2 \\
                                       a_3 \\
                                     \end{array}
                                   \right)
 =\left(
    \begin{array}{ccc}
      0 & -vJ_0(A_1/\omega) & 0 \\
      -vJ_0(A_1/\omega) & -i\alpha_2 & -vJ_0(A_2/\omega) \\
      0 & -vJ_0(A_2/\omega) & 0 \\
    \end{array}
  \right)\left(
           \begin{array}{c}
             a_1 \\
             a_2 \\
             a_3 \\
           \end{array}
         \right).
\end{align}

As is well
known, the periodic time-dependent equation \eqref{c_n} admits solutions in the form of Floquet states
$(c_1,c_2,...,c_N) =(c'_1,c'_2,...,c'_N)\exp(-i \varepsilon t)$, where $\varepsilon$ is the quasienergy and the amplitudes $(c'_1,c'_2,...,c'_N)$
are periodic with the driving period $T=2\pi/\omega$. Applying the well-established Floquet theorem to the periodic
system \eqref{Model for c}, we can construct the approximate Floquet solutions $c_n =c'_n(t)\exp(-i \varepsilon t)=a_n\exp[-i\int \epsilon_n (t)dt]\simeq
a'_n\exp[-i\int \epsilon_n (t)dt-iEt]$ with $c'_n(t)\simeq a'_n\exp[-i\int \epsilon_n (t)dt], \varepsilon\simeq E, a_n=a'_n\exp(-iEt), n=1,2,3$,
where  constants $a'_n$ and $E$ are the eigenvector
components and the eigenvalue of the time-independent
version of equation \eqref{Model for a1} respectively. Inserting such a form of $a_n=a'_n\exp(-iEt)$
 into  equation \eqref{Model for a1}, we obtain the eigenvalues (approximate quasienergies)
\begin{equation}\label{eigvalue}
  \varepsilon_1=0,~~\varepsilon_{2,3}=\frac12(-i\alpha_2\pm\Theta),
\end{equation}
and  the corresponding Floquet modes
 \begin{align}\label{ut}
  |u_1(t)\rangle&=-J_0(A_2/\omega)e^{-i\frac{A_1}{\omega}\sin(\omega t)}|1\rangle+0|2\rangle+
J_0(A_1/\omega)e^{-i\frac{A_2}{\omega}\sin(\omega t)}|3\rangle,\nonumber\\
  |u_{2,3}(t)\rangle&=J_0(A_1/\omega)e^{-i\frac{A_1}{\omega}\sin(\omega t)}|1\rangle+\frac12(i\alpha_2\mp\Theta)|2\rangle\nonumber\\
  &+
J_0(A_2/\omega)e^{-i\frac{A_2}{\omega}\sin(\omega t)}|3\rangle,
\end{align}
where $\Theta=\sqrt{\gamma^2-\alpha_2^2}$ with $\gamma^2=4v^2[J_0^2(A_1/\omega)+J_0^2(A_2/\omega)]$.
The Floquet mode $|u_1(t)\rangle$ corresponds to  a so-called
dark Floquet state in the dissipative system, which seemingly has zero quasienergy and zero population at the
intermediate state $|2\rangle$.

At time $t$, the wave function evolves according to
\begin{equation}\label{psit2}
  |\psi(t)\rangle=\sum_{n=1}^{3}F_ne^{-i\varepsilon_nt}|u_n(t)\rangle,
\end{equation}
where $F_n$ are superposition coefficients determined by the initial states, which can be calculated as
\begin{equation}\label{F-c0}
  \left(
    \begin{array}{c}
      F_1 \\
      F_2 \\
      F_3 \\
    \end{array}
  \right)=T^{-1}\left(
    \begin{array}{c}
      c_1(0)\\
      c_2(0)\\
      c_3(0)\\
    \end{array}
  \right)
\end{equation}

\begin{widetext}
\begin{equation}\label{T-1}
 T^{-1}= \frac{1}{|T|}\left(
    \begin{array}{ccc}
      -\Theta J_0(A_2/\omega) & 0 & \Theta J_0(A_1/\omega) \\
      \frac12(i\alpha_2+\Theta)J_0(A_1/\omega) & -J_0^2(A_1/\omega)+J_0^2(A_2/\omega) & \frac12(i\alpha_2+\Theta)J_0(A_2/\omega) \\
      \frac12(i\alpha_2-\Theta)J_0(A_1/\omega) & J_0^2(A_1/\omega)+J_0^2(A_2/\omega) & -\frac12(i\alpha_2-\Theta)J_0(A_2/\omega) \\
    \end{array}
  \right).
\end{equation}
\end{widetext}
Here $|T|=\Theta[J_0^2(A_1/\omega)+J_0^2(A_2/\omega)]$.

 As an example, throughout our paper, we only consider the case in which the particle is initially
prepared in the site 1. Applying the initial
condition $c_1(0) = 1, c_2(0) = 0,c_3(0) = 0$ to Eqs.~\eqref{psit2} and \eqref{F-c0}
yields the analytical solutions of Eq.~\eqref{Model for c}  in the forms
\begin{align}\label{ct}
  c_1(t) & =e^{-i\frac{A_1}{\omega}\sin(\omega t)}\Big[\frac{J_0^2(A_2/\omega)}{J_0^2(A_1/\omega)
  +J_0^2(A_2/\omega)}e^{-i\varepsilon_1t}+\frac12f_+(t)J_0^2(A_1/\omega)\Big]\nonumber\\
  c_2(t)&=\frac{J_0(A_1/\omega)(\alpha_2^2+\Theta^2)}{4|T|}\left(e^{-i\varepsilon_3 t}-e^{-i\varepsilon_2 t}\right)\nonumber\\
   c_3(t) & =e^{-i\frac{A_2}{\omega}\sin(\omega t)}J_0(A_1/\omega)J_0(A_2/\omega)\nonumber\\
   &\times\left(-\frac{e^{-i\varepsilon_1t}}{J_0^2(A_1/\omega)+J_0^2(A_2/\omega)}
  +\frac12f_+(t)\right),
\end{align}
where
\begin{align}\label{ft}
  f_+(t)&=\frac{i\alpha_2+\Theta}{|T|}e^{-i\varepsilon_2t}
  -\frac{i\alpha_2-\Theta}{|T|}e^{-i\varepsilon_3t}\nonumber\\
  &=\frac{1}{|T|}e^{-\frac12\alpha_2 t}\left(i\alpha_2(e^{-\frac12 i\Theta t}-e^{\frac12 i\Theta t})
  +\Theta(e^{-\frac12 i\Theta t}+e^{\frac12 i\Theta t})\right).
\end{align}
The analytical solutions of equation \eqref{Model for c} incorporate all
available time-dependent information about the system.
Specifically, the probability $P_n(t)$ of finding the system to be in
state $|n\rangle$ at time $t$ (here also termed the population of site $n$)
is the absolute square of the occupation amplitude in each local state
\begin{align}\label{Pn}
 P_n(t)=|c_n|^2,
\end{align}
and the sum gives the total population (probability),
\begin{align}\label{Pt}
 P(t)=\sum_{n}|c_n|^2.
\end{align}

Substituting \eqref{eigvalue} into \eqref{ct} and \eqref{Pn}, we have
\begin{align}
  P_1(t)&=\Big|\frac{J_0^2(A_2/\omega)}{J_0^2(A_1/\omega)+J_0^2(A_2/\omega)}+
  \frac12 f_+(t)J_0^2(A_1/\omega)\Big|^2, \label{p1} \\
    P_2(t)&=\frac{4J_0^2(A_1/\omega)}{|\Theta|^2}e^{-\alpha_2 t}\left\{\begin{array}{cc}
                                                                           \sin^2\frac12 |\Theta| t, & \alpha_2<\gamma, \\
                                                                           \sinh^2\frac12 |\Theta| t, & \alpha_2>\gamma,
                                                                         \end{array}\right.\label{p2}\\
    P_3(t)&=\Big|-\frac{J_0(A_1/\omega)J_0(A_2/\omega)}{J_0^2(A_1/\omega)+J_0^2(A_2/\omega)}+
  \frac12 f_+(t)J_0(A_1/\omega)J_0(A_2/\omega)\Big|^2.\label{p3}
\end{align}

As can be observed from the above expression, the solutions \eqref{p1}-\eqref{p3}
are distinguished as two classes: underdamped or
overdamped, relying on whether the value of $\Theta$ is taken of a real or complex number.
When the loss coefficient is small, $\alpha_2<\gamma$, then $\Theta$ is a positive
and real
number, $\Theta=|\Theta|$, and thus the function $f_+$ in the solutions \eqref{p1}-\eqref{p3} is of the form
\begin{align}\label{ftRed1}
  f_+(t)=\frac{2}{J_0^2(A_1/\omega)+J_0^2(A_2/\omega)}e^{-\frac12\alpha_2 t}\frac{\sin(\frac12 |\Theta| t+\beta)}{\sin\beta},
\end{align}
where $\cos\beta=\alpha_2/\gamma$. In this regime, the solutions exhibit
underdamped oscillations, which occur at the frequency of $|\Theta|/2$  but with
a decaying amplitude  proportional to the
exponential $\exp(-\alpha_2t/2)$, and an increase of the loss coefficient $\alpha_2$ will increase the rate at which
the total probability decays to
approach the equilibrium value. When the loss coefficient is large, $\alpha_2>\gamma$, then $\Theta$  becomes a
purely imaginary number, $\Theta=i|\Theta|$, and the function $f_+$ is expressible
using hyperbolic sines  as
\begin{align}\label{ftRed2}
  f_+(t)=\frac{2}{J_0^2(A_1/\omega)+J_0^2(A_2/\omega)}e^{-\frac12\alpha_2 t}\frac{\sinh(\frac12 |\Theta| t+\beta')}{\sinh\beta'},
\end{align}
 where $\cosh\beta'=\alpha_2/\gamma$.
 In this case, the system
enters the overdamping regime, where the solutions evolve in the exponential form of $\exp[(-\alpha_2\pm|\Theta|)t/2]$, and the decaying
term  $\exp(-\alpha_2t/2)$  will be partly compensated by
the monotonic increasing hyperbolic sine function, thereby enabling the system to
 access the equilibrium  much more slowly than that predicted by expression \eqref{ftRed1}. In the overdamped regime, a  surprising result is that an increase in the loss
 coefficient will produce a slower decay rate of the total probability. Equation \eqref{p1}-\eqref{p3} also shows that in the limit of
indefinitely large loss coefficient (thus $|\Theta|\rightarrow \alpha_2$), the initial-state population tends to unity, $P_1{(\alpha_2\rightarrow \infty)}\rightarrow 1$, and the particle will remain frozen
in its initially occupied lossless site for long times. This behavior seems reminiscent of the effect of loss-induced localization
studied in Reference \cite{Li}, and the Zeno-like effect addressed in other literatures\cite{Barontini,Shchesnovich1,Witthaut2, Hofstetter,Ott2}.
The boundary between underdamped oscillation and overdamping  occurs when $\alpha_2=\gamma$, which is called critical damping.
When critical damping occurs, the functions of \eqref{ftRed1} and \eqref{ftRed2} coalesce to the same form, and the system evolves in time toward the  equilibrium in the form of exponential $\exp(-\alpha_2t/2)$. From the expression of $\gamma^2=4v^2[J_0^2(A_1/\omega)+J_0^2(A_2/\omega)]$, it can be apparently seen that the critical value of $\alpha_2=\gamma$ can be tuned by adjusting the driving parameters, which provides a fascinating avenue for manipulation of the transition from underdamped oscillation to overdamping.

In addition, we  are particularly interested in  the asymptotic behavior of
the populations $(P_n)_{\textmd{asy}}\equiv P_n(t\rightarrow\infty)$ and $P_{\textmd{asy}}\equiv\sum_{n}P_n(t\rightarrow\infty)$ at $t\rightarrow\infty$.
According to the results of high-frequency Floquet
analysis, in the limit of $t\rightarrow\infty$, we have $e^{-i\varepsilon_{2,3}t}\to0$ and $f_+(t)\to0$; for the given initial state $c_1(0) = 1, c_2(0) = 0,c_3(0) = 0$, the wave function $|\psi(t)\rangle$ will be asymptotically driven to a sink state,
\begin{equation}\label{sink}
|\psi(t\rightarrow\infty)\rangle =F_1e^{-i\varepsilon_1 t} \left(
    \begin{array}{c}
      -J_0(A_2/\omega)e^{-i\frac{A_1}{\omega}\sin(\omega t)} \\
      0 \\
      J_0(A_1/\omega)e^{-i\frac{A_2}{\omega}\sin(\omega t)} \\
    \end{array}
  \right)
\end{equation}
with $\varepsilon_1=0$, $F_1=-J_0(A_2/\omega)/[J_0^2(A_1/\omega)+J_0^2(A_2/\omega)]$, and the associated norms are given by
\begin{align}\label{ctlimit1}
  (P_1)_{\textmd{asy}}&=\frac{J_0^4(A_2/\omega)}{[J_0^2(A_1/\omega)+J_0^2(A_2/\omega)]^2}, \\
  (P_2)_{\textmd{asy}}&=0,\\
 (P_3)_{\textmd{asy}}&=\frac{J_0^2(A_1/\omega)J_0^2(A_2/\omega)}
 {[J_0^2(A_1/\omega)+J_0^2(A_2/\omega)]^2},\\
 P_{\textmd{asy}}&=\frac{J_0^2(A_2/\omega)}{J_0^2(A_1/\omega)+J_0^2(A_2/\omega)}.\label{ctlimit4}
\end{align}
As indicated in Equation \eqref{sink}, the sink state is just the so-called dark Floquet state,
which seemingly has zero quasienergy and zero population at the
intermediate state $|2\rangle$. As the dark Floquet state is reached
by dissipative dynamics, the lossy site is not
excited and the system is  projected onto the loss-free
 dynamics. From equations \eqref{ctlimit1}-\eqref{ctlimit4}, we readily observe the following
circumstances.

\emph{Case I}: If $A_2=0$, $A_1$ is taken arbitrarily, that is, the periodic driving
applied only to the left-end site, then
\begin{equation}\label{c1tI}
  (P_1)_{\textmd{asy}}=\frac{1}{[1+J_0^2(A_1/\omega)]^2}\geq\frac14,
\end{equation}
\begin{equation}\label{c3tI}
  (P_3)_{\textmd{asy}}=\frac{J_0^2(A_1/\omega)}{[1+J_0^2(A_1/\omega)]^2}\leq\frac14,
\end{equation}
and the total population
\begin{equation}\label{totaI}
   P_{\textmd{asy}}=\frac{1}{1+J_0^2(A_1/\omega)}\geq\frac12.
\end{equation}
In this case, the ratio of  asymptotic probability $(P_1)_{\textmd{asy}}$ to $(P_3)_{\textmd{asy}}$ is given by $[|c_1(t)|^2/|c_3(t)|^2](t\to+\infty)=1/J_0^2(A_1/\omega)\geq1$. From equation \eqref{totaI}, it follows that the periodic driving, applied only to  the left-end site, may lead to a enhanced value of  $P_{\textmd{asy}}$ and thus to a reduced loss of the total population, as compared with the undriven case where $P_{\textmd{asy}}=1/2$.
This important finding can be exploited for improvement of
the antileakage capability in the dissipative system.
Specifically, if we choose $A_1/\omega$ to be the zeros of the Bessel function $J_0(A_1/\omega)$, then we obtain
$(P_1)_{\textmd{asy}}=1$, $(P_3)_{\textmd{asy}}=0$ and $P_{\textmd{asy}}=1$.
In such a case, it seems that the particle remains fixed
in its initial site and the system is effectively lossless, which recovers the CDT effect.

\emph{Case II}: If $A_1=0$, $A_2$ is arbitrary, that is, the periodic driving applied only to  the right-end site, then we have
\begin{equation}\label{c1tII}
  (P_1)_{\textmd{asy}}=\frac{J_0^4(A_2/\omega)}{[1+J_0^2(A_2/\omega)]^2}\leq\frac14,
\end{equation}
\begin{equation}\label{c3tII}
  (P_3)_{\textmd{asy}}=\frac{J_0^2(A_2/\omega)}{[1+J_0^2(A_2/\omega)]^2}\leq\frac14,
\end{equation}
and the total population
\begin{equation}\label{totaII}
   P_{\textmd{asy}}=\frac{J_0^2(A_2/\omega)}{1+J_0^2(A_2/\omega)}\leq\frac12.
\end{equation}
Contrary to \emph{case I}, the periodic driving, applied merely to the right-end site, instead makes the ratio of
$(P_1/P_3)_{\textmd{asy}}$ to be small than 1, that is,  $[|c_1(t)|^2/|c_3(t)|^2(t\to+\infty)]=J_0^2(A_2/\omega)\leq1$, and produces a greater total population loss than the undriven counterpart. This case does not lie in the focus of the following study.

\emph{Case III}: If $A_1=A_2\neq0$, that is, the periodic driving applied equally to  the two end sites, then we have
$(P_1)_{\textmd{asy}}=|c_1(t)|^2(t\to+\infty)=\frac14$, $(P_3)_{\textmd{asy}}=|c_3(t)|^2(t\to+\infty)=\frac14$ and $P_{\textmd{asy}}=P(t\to+\infty)=\frac12$.
Under this circumstance, the asymptotic evolution at $t\to+\infty$ seems to behave exactly the same as the undriven case.

Following the high-frequency approximation analysis mentioned above, we summarize
the main results as follows: (i)
the periodic driving can be utilized not only to manipulate the damping form from underdamped oscillation to overdamping, but also to reduce the decay of the total population in the dissipative three-site system;
(ii) the sink state (dark Floquet state), to which the system is asymptotically driven, has unique population distribution which depends not on the strength of loss coefficient but only on  the driving parameters.

\begin{figure}[htp]
\center
\includegraphics[width=8cm]{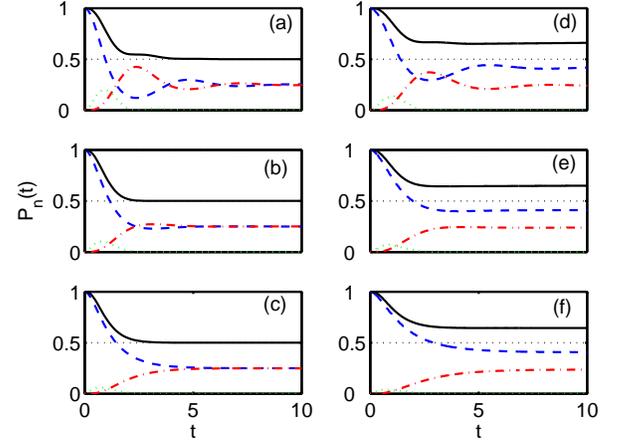}
\caption{(color online) Dynamics of the three-site system
for the undriven ($A_1=0, A_2=0$, left column) and driven ($A_1=20, A_2=0$, right column) cases
 with
 $\omega=20, v = 1$ and different values of $\alpha_2$. From top to bottom: (a) and (d) $\alpha_2=1$;
(b) and (e) $\alpha_2=2$; (c) and (f) $\alpha_2=3$. The particle is initialized in the first site. Shown here are
the numerical probabilities $P_1 = |c_1|^2$ (blue dashed line), $P_2 = |c_2|^2$ (green dotted line), $P_3 = |c_3|^2$ (red dashed-dotted line), and the total  probability $P(t)=\sum_{n=1}^3P_n$ (black solid line).} \label{fig2}
\end{figure}

To verify the above analytical arguments,  we solve the
 equation \eqref{Model for c} numerically
with the particle initially localized in the first site. The time evolutions of all probabilities with increasing values of loss coefficient for the undriven and driven cases are shown in Fig.~\ref{fig2} (a-c) and Fig.~\ref{fig2} (d-f) respectively, in both of which  different types of damping are  possible as we have predicted. Fig.~\ref{fig2} (a-c) illustrates representative examples of the undriven model,
for the three regimes of damping: underdamped oscillation ($\alpha_2=1$), critical
damping ($\alpha_2=2$) and overdamping ($\alpha_2=3$). In the undriven model, the total population
will decrease to half of its initial value, independently of the  values of loss coefficient.
 By contrast, as exhibited in Fig.~\ref{fig2} (d-f), for the driven model with the periodic driving field acting only on the first site, the system still undergoes the transition from underdamped oscillation to overdamping as the loss coefficient gets stronger ($\alpha_2=1,2,3$ from top to bottom), but the total probability approaches
the same  equilibrium value (independent of $\alpha_2$) after a period of time, which nevertheless becomes higher as compared with the undriven counterpart (the value of $0.5$).

\begin{figure}[htp]
\center
\includegraphics[width=8cm]{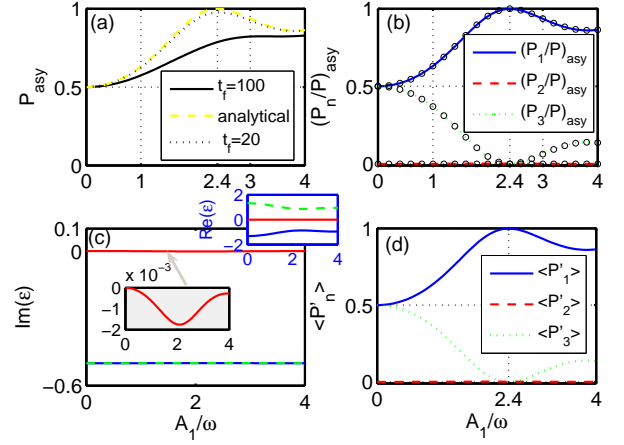}
\caption{(color online) Three-site model \eqref{Model for c} with $A_2=0$. (a) Comparison between
analytical and numerical  results of asymptotic (equilibrium) value of the total probability versus the driving parameter $A_1/\omega$.  The analytical results $P_{\textmd{asy}}= P(t\rightarrow\infty)$
are given by the formula \eqref{totaI}, and the numerical correspondences $\langle P\rangle_{\rm{equ}}=\frac{1}{\Delta}\int_{t_f-\Delta}^{t_f}P(t)dt $ are obtained from
the original three-site system \eqref{Model for c} with different integration times $t_f$. (b) $(P_n/P)_{\rm{asy}}$, given by the analytical formulas \eqref{ctlimit1}-\eqref{ctlimit4}, versus $A_1/\omega$. Circles are for numerical correspondences $\langle P_n/P\rangle_{\rm{equ}}$ with $t_f=100$. (c) Imaginary and real ( upper-right corner) parts of numerical quasienergies versus $A_1/\omega$ for the three-site system \eqref{Model for c}. The inset of (c) shows an enlargement of imaginary  part of the zero-quasienergy (red line). (d) The time-averaged probability distribution
of the dark Floquet state corresponding to a nearly zero quasienergy in panel (c).
In all panels, we only consider the case the periodic driving field is only applied to site 1. The other parameters and initial condition  are the same as those in Fig.~\ref{fig2} (d): $A_2=0, \omega=20,v=1,\alpha_2=1$, and $ (c_1(0),c_2(0),c_3(0))=(1,0,0)$. The averaging time used in (a-b) is $\Delta=t_f/2$.} \label{fig3}
\end{figure}

In fact, the real integration time cannot be infinite for the time integration of the equation \eqref{c_n}. For comparison, we define the numerical correspondence of
asymptotic (equilibrium) value of all probabilities as $\langle P_n\rangle_{\rm{equ}}=\frac{1}{\Delta}\int_{t_f-\Delta}^{t_f}P_n(t)dt $ and $\langle P\rangle_{\rm{equ}}=\sum_n\langle P_n\rangle_{\rm{equ}}$ with appropriate averaging time interval $\Delta$. In the late calculations, we will take $\Delta=t_f/2$ to ensure that the initial non-steady process is omitted. In Fig.~\ref{fig3} (a), we
have shown the analytical results of $P_{\rm{asy}}$ (yellow dashed line) given by \eqref{totaI} versus the driving parameters
$A_1/\omega$  with  other parameters as in Fig.~\ref{fig2} (d), and found that as $A_1/\omega$ is increased from zero, the value of $P_{\rm{asy}}$ grows from $0.5$ and reaches its maximum (of 1) at $A_1/\omega=2.4$, the zero of $J_0(A_1/\omega)$.
We also compare the analytical results with the numerical correspondences calculated
via integration of equation \eqref{Model for c} with different integration
times. As shown in Fig.~\ref{fig3} (a), the numerical correspondences  with $t_f=20$
are in good agreement with the analytical results, whereas the numerical results (remain to be greater than or equal to $0.5$) are principally  below the  analytical ones if we extend the integration time to $t_f=100$. Another important observation is that
the numerical results of ratio $\langle P_n/P \rangle_{\rm{equ}}$ (circles) obtained with $t_f=100$ (or even indefinitely longer), nevertheless, fit quite well to the analytical results of $(P_n/P)_{\rm{asy}}$, as seen in Fig.~\ref{fig3} (b).
To gain more insight into this problem,  we have  numerically computed the quasienergies and Floquet
states of the original system model \eqref{Model for c}.
The numerically computed quasienergies are shown in Fig.~\ref{fig3} (c). It is clear from Fig.~\ref{fig3} (c) that there exists a
dark Floquet state whose quasienergy (red line) seems to be zero for all of the values of $A_1/\omega$. Notice however,
as shown in the inset, that the zero quasienergy spectrum is in
fact a complex quasienergy spectrum with small negative imaginary parts.
This dark Floquet state stands out not only for its nearly zero quasienergy
but also for its unique population distribution among the local sites.
We display the time-averaged population $ \langle P'_n\rangle=[\int_0^Tdt|c'_n|^2]/T$
corresponding to the dark Floquet state in Fig.~\ref{fig3} (d). Remarkably, the time-averaged population distribution
of dark Floquet state seems to be equivalent to the behaviors of ratio $(P_n/P)_{\rm{asy}}$ shown in Fig.~\ref{fig3} (b), and accordingly, the dark Floquet state has negligible
population at site $2$.

The above numerical simulations reveal that for the three-site  system \eqref{Model for c}, driven by the localized dissipation at the central site and by the periodic  field acting only on the left-end site, the short-time evolution of all probabilities obtained from
the original model \eqref{Model for c} agrees well with the analytical results given by the high-frequency approximation. However, a significant deviation between them emerges in the long-time evolution. Interestingly,  the numerical results of ratio $\langle P_n/P \rangle_{\rm{equ}}$  at any long-enough evolution time are well consistent with the analytical $(P_n/P)_{\rm{asy}}$ at $t\rightarrow\infty$ and the time-averaged population $\langle P'_n\rangle$ corresponding to the dark Floquet state as well. The underlying physics can be understood by noting that the sink state (dark Floquet state) itself will continue to decay very slowly as $\exp[\rm{Im}(\varepsilon_1)]$, with $\rm{Im}(\varepsilon_1)$ being non-zero but a  small negative number. The transient nature of sink (equilibrium)  state is quite natural, since the losses are not balanced by any gain, and apparently, the very small negative imaginary part of quasienergy would be intrinsically related to a small nonzero population at the lossy site $2$, which can be numerically confirmed by amplified examination (not shown) of the population distribution of the associated dark Floquet state.
In practice, the application of periodic driving field merely to the left-end site offers an efficient and feasible tool to increase the equilibrium value of total probability, and thus to attenuate the effective decay in the dissipative three-site system  for a certain long interval of time.  The foreseeable disadvantage for this scheme, however, is that the probabilities approach zero and all probabilities will be lost in the long run. This transient nature presents
a major obstacle for practical
applications in the control of the dynamics of an open quantum system  with localized dissipation.
Finally, it is worth emphasizing again, in the dissipative system, the fact that the quasienergy of the dark Floquet state has a very small non-zero imaginary part will produce a nontrivial physical effect, which is extremely different from the conservative system, where the nonzero but vanishingly small all-real quasienergy  of dark Floquet state accumulates no physical effect in long-time evolution of quantum states.

\begin{figure}[htp]
\center
\includegraphics[width=8cm]{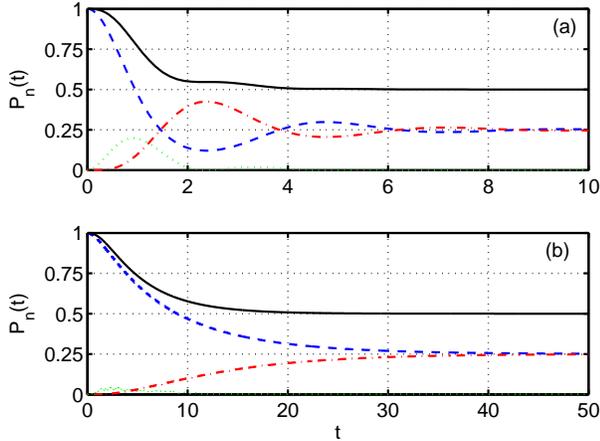}
\caption{(color online) Time evolutions of the probabilities $P_n = |c_n |^2 ( n= 1, 2, 3)$ in the three-site system \eqref{Model for c} for the particle initially occupying the first site.
(a) $A_1=0, A_2=0$; (b) $A_1=40, A_2=40$. The other
parameters are set as $\omega=20,v=1,\alpha_2=1$. Shown here are the numerical probabilities $P_1(t)$ (blue dashed line), $P_2(t)$ (green dotted line), $P_3(t)$ (red dashed-dotted line), and the total  probability $P(t)$ (black solid line).} \label{fig4}
\end{figure}

Fig.~\ref{fig4} illustrates a representative example of tuning the transition from underdamped oscillation to overdamping by the periodic driving field. For the undriven case, $A_1=A_2=0$, we find that both $P_1$ and $P_3$ exhibit oscillations in the initial time interval, and then tend to a constant of $0.25$, which signals a form of underdamped oscillation, as shown in Fig.~\ref{fig4} (a). As the periodic field is applied equally to the two end sites, $A_1=A_2\neq0$, both $P_1$ and $P_3$ have no oscillations, and only exhibit a monotonic approach toward the equilibrium that is exactly the same as  the undriven counterpart, as shown in Fig.~\ref{fig4} (b). This behavior of the driven case is a manifestation of overdamping.
These numerical results demonstrate that the transition between underdamped oscillation and overdamping
can be controlled with high precision by adjusting the driving parameters.

\subsection{four-site system}
We now turn to the case of the four-site system
where the dynamical
equations are
\begin{align}\label{Model for c4}
 i\frac{\partial}{\partial t}\left(
                                     \begin{array}{c}
                                       c_1 \\
                                       c_2 \\
                                       c_3 \\
                                       c_4 \\
                                     \end{array}
                                   \right)
 =\left(
    \begin{array}{cccc}
      A_1\cos(\omega t) & -v & 0 & 0\\
      -v & -i\alpha_2 & -v & 0 \\
      0 & -v & 0 & -v \\
      0 & 0 & -v & A_2\cos(\omega t) \\
    \end{array}
  \right)\left(
           \begin{array}{c}
             c_1 \\
             c_2 \\
             c_3 \\
             c_4 \\
           \end{array}
         \right).
\end{align}

\begin{figure}[htp]
\center
\includegraphics[width=8cm]{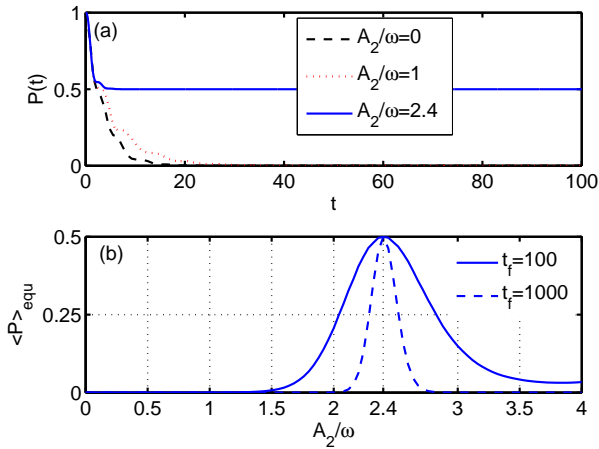}
\caption{(color online) Four-site system described by equation \eqref{Model for c4}. (a) Time evolutions of the total probability $P(t)$ under
various driving conditions. (b) Numerical value of $\langle P \rangle_{\rm{equ}}$ as a function of driving parameter $A_2/\omega$, with two different integration times $t_f=100$ and $t_f=1000$. In both panels, we start the system with the particle at site 1, and set other parameters as $A_1=0, \omega=20,v=1,\alpha_2=1$.} \label{fig5}
\end{figure}

As far as the four-site system ($N$ is even) is concerned,
the focus of our studies is on the situation where the periodic driving field
is applied only to the right-end site. We plot in Fig.~\ref{fig5}(a) the time evolutions of the total population by direct numerical integration of the Schr\"{o}dinger
equation \eqref{Model for c4} with $A_1 = 0$. We start a particle at site 1 and
fix the loss coefficient as $\alpha_2=1$. The results are presented in Fig.~\ref{fig5}(a) for three typical driving conditions, where the total probabilities  are found to decay from
1 to 0 rapidly for both cases of $A_2/\omega=0$ and $A_2/\omega=1$. For $A_2/\omega=2.4$, however,
 we observe a quite different behavior: the total probability decays to reach a equilibrium
value of $0.5$, showing a remarkable coincidence with the counterpart of undriven three-site system.
The basic explanation
of this  phenomenon is as follows. If the driving parameters $A_2/\omega$ are
tuned to satisfy the zeroth-order Bessel function $J_0(A_2/\omega)=0$, CDT occurs between the driven right-end site and the undriven lattice sites, and therefore the system behaves like an undriven three-site system with localized dissipation at site 2. Thus, the coaction of CDT effect and dark state (existing in the three-site system) will give origin to the phenomenon that the dissipative four-site system evolves in time toward a steady state which keeps the total probability at a constant level of $0.5$.
Such a suppression of the total probability loss is more clearly
demonstrated in Fig.~\ref{fig5} (b), where the numerical equilibrium $\langle P \rangle_{\rm{equ}}$ versus the driving parameter $A_2/\omega$ is presented with two different evolution times. It is observed that there exist the
peaks in this quantity $\langle P \rangle_{\rm{equ}}$ centered on $A_2/\omega=2.4$, the
zero of $J_0(A_2/\omega)$, for both evolution times. As the evolution time is extended, the peak becomes narrower, but the maximum value of $\langle P \rangle_{\rm{equ}}$ remains unchanged nevertheless.

\begin{figure}[htp]
\center
\includegraphics[width=8cm]{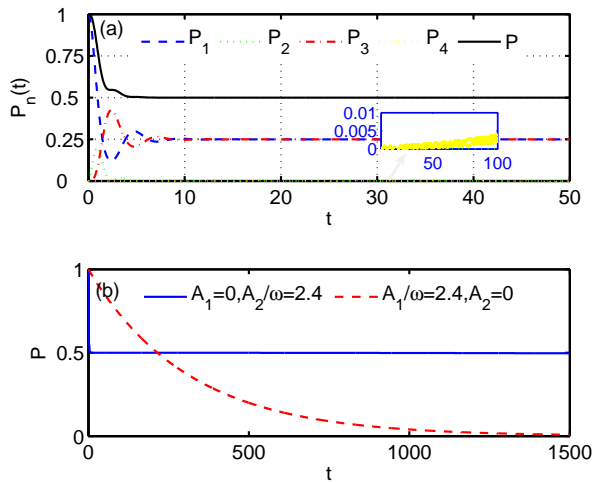}
\caption{(color online) Four-site system \eqref{Model for c4} for the particle initially occupying the first site. (a) Time evolutions of all probabilities, with $A_1=0, A_2/\omega=1$ and the same parameters as in Fig.~\ref{fig5}. Inset shows amplification of the time evolutions of $P_2$ and $P_4$. (b) Time evolutions of total probabilities for two cases of $A_1=0, A_2/\omega=2.4$ and $A_1/\omega=2.4, A_1=0$, with $\omega=20,v=1,\alpha_2=1$. } \label{fig6}
\end{figure}

To better understand the phenomenon presented in Fig.~\ref{fig5}, we carefully investigate the time-dependent behaviors of all probabilities, as illustrated in  Fig.~\ref{fig6} (a). The inset of Fig.~\ref{fig6} (a) gives an enlargement of the time evolutions of probabilities  $P_2$ and $P_4$, where we observe that the two quantities keep very small values over the evolution time but the probability $P_4$ is nonetheless more appreciable.
The exceedingly small and less  appreciable value of  time-dependent probability $P_2$ implies that the equilibrium value will keep good stability over long-enough evolution time. This implication is  collaborated  by
our numerical simulation (not shown) if we extend the integration time to a larger order of magnitude. It might be taken for granted that due to the occurrence of CDT, application of the periodic driving field solely to the first site will lead to a decoupling of the site 1 and the lossy site 2 (hence to prevention of the particle from reaching the leaking site), therefore protecting the system from dissipation. However, the fact is not so simple as is thought. In Fig.~\ref{fig6}, we give the direct comparison of the time-dependent total probabilities between the case of $A_1=0, A_2/\omega=2.4$ and
  $A_2=0, A_1/\omega=2.4$. We easily observe from Fig.~\ref{fig6} that for the case of  $A_2=0, A_1/\omega=2.4$ (here the periodic driving is applied only to the left-end site, and approximate CDT occurs between  site 1 and the undriven lattice sites), the total probability slowly falls from its initial value to zero as the time is increased, whereas for the case of  $A_1=0, A_2/\omega=2.4$ (the periodic driving is applied only to the the right-end site,  and approximate CDT occurs between  the right-end site and other undriven lattice sites), the total probability decays to approach a steady value with long-term stability.

\subsection{system with other  number of
sites}
Our analysis above is given for three- and four-site systems,
but similar behavior can be obtained  for other lattice systems
as well.

\begin{figure}[htp]
\center
\includegraphics[width=8cm]{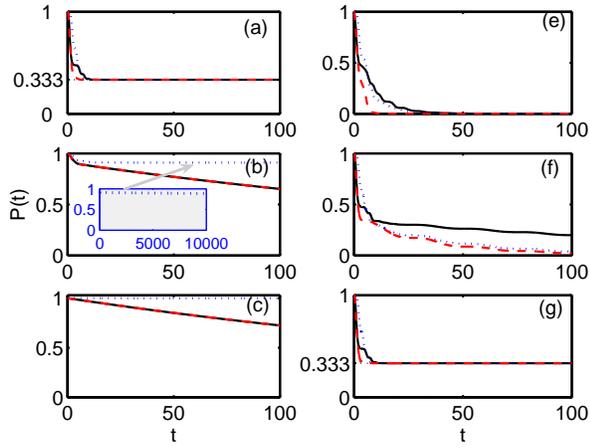}
\caption{(color online) Time evolutions of the total probability for the 5-site system with $A_2=0$ (left column) and the 6-site system with $A_1=0$ (right column), starting from
 the particle at site 1.  From top to bottom on the left: (a) $A_1/\omega=0$;
(b) $A_1/\omega=2$; (c) $A_1/\omega=2.4$; From top to bottom on the right: (d) $A_2/\omega=0$;
(e) $A_2/\omega=2$; (f) $A_2/\omega=2.4$. Other parameters are $\omega=20,v=1,\alpha_2=1$.
 Shown here are
the numerical results for the parameter set of loss coefficient $(\alpha_2,\alpha_4)=(0,1)$ (blue dotted line),
$(\alpha_2,\alpha_4)=(1,1)$ (red dashed line), and $(\alpha_2,\alpha_4)=(1,0)$ (black solid line). The inset of panel (b) illustrates the long-time evolution of the total probability in the 5-site system  for the parameter set $(\alpha_2,\alpha_4)=(0,1)$, with $A_2=0, A_1/\omega=2$.} \label{fig7}
\end{figure}

The dissipative dynamics of the driven $N$-site systems
is elaborated by directly integrating the time-dependent
Schr\"{o}dinger equation \eqref{c_n} with the particle initially localized
at site 1. Fig.~\ref{fig7} shows some examples of the dynamics for other finite number of
sites with
different driving parameters and with different values of loss coefficient. The left column shows the dynamics for $N=5$ with the periodic driving field applied only to the left-end site and  the right column for $N=6$ with the periodic driving field  only to the right-end site. For the system with $N=5$, it can be seen that (i) when the periodic driving is not presented, that is, $A_1/\omega=0$, the total probabilities $P(t)$ rapidly decay to the same equilibrium
value of $0.3333$ for all values of loss coefficient; (ii) as the periodic driving is switched on, for both parameter sets of loss coefficient
$(\alpha_2,\alpha_4)=(1,1)$ and $(\alpha_2,\alpha_4)=(1,0)$ (where the parameter sets both correspond to the same case in which the first lossy site from the left is site 2), the probabilities exhibit the same behavior with persistently slow decay though the total number of lossy sites for both are distinct; while for the  parameter set $(\alpha_2,\alpha_4)=(0,1)$ (i.e., the case that the first lossy site from the left is site 4), the loss of total probabilities can be greatly reduced with respect to the undriven case and the enhanced equilibrium
value of total probabilities can be kept at a stable level even though total evolution
time is extended to a much larger order of magnitude (see the inset of Fig.~\ref{fig7}(b)).
Notice that at $A_1/\omega=2.4$, the zero of $J_0(A_1/\omega)$, the total probability $P(t)$ remains near unity over the evolution time for the parameter set $(\alpha_2,\alpha_4)=(0,1)$, which means a vanishing decay of total population of the quantum wave packet.
For the system with $N=6$, on the other hand, we
observe that the total probabilities $P(t)$ tend rapidly to zero
for both cases of $A_2/\omega=0$ and $A_2/\omega=2$.
However, it is interesting to note that at $A_2/\omega=2.4$, the zero of $J_0(A_2/\omega)$,
the total probabilities $P(t)$ decay to approach the steady value of $0.3333$ for any set of  loss coefficient, which works in analogy to the undriven five-site system (one can see by comparing Fig.~\ref{fig7} (g) and (a)).

\begin{figure}[htp]
\center
\includegraphics[width=8cm]{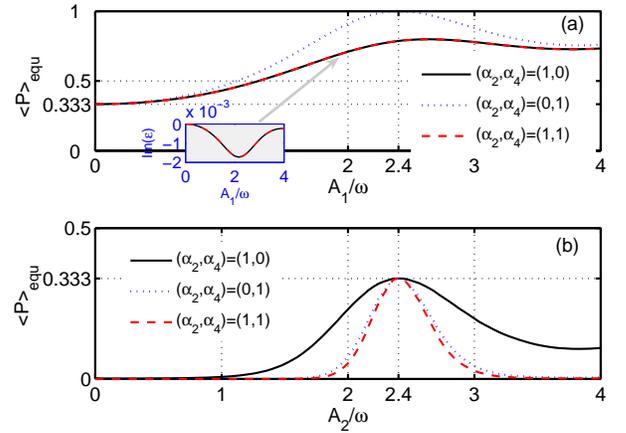}
\caption{(color online) (a) Numerical value of $\langle P \rangle_{\rm{equ}}$ as a function of driving parameter $A_1/\omega$ for the 5-site system with $A_2=0$ for different parameter sets of loss coefficient. Other parameters are the same as the ones in Fig.~\ref{fig7} (a-c). (b)
$\langle P \rangle_{\rm{equ}}$ versus $A_2/\omega$ for the 6-site system with $A_1=0$ for different parameter sets of loss coefficient. Other parameters are the same as the ones in Fig.~\ref{fig7} (d-f). In both panels, we assume that the system is initialized with the particle at site 1.
The inset of panel (a) gives the ampliation of  imaginary part of nearly zero quasienergy corresponding to the dark Floquet state for both parameter sets  $(\alpha_2,\alpha_4)=(1,1)$ and $(\alpha_2,\alpha_4)=(1,0)$.} \label{fig8}
\end{figure}

\begin{figure}[htp]
\center
\includegraphics[width=8cm]{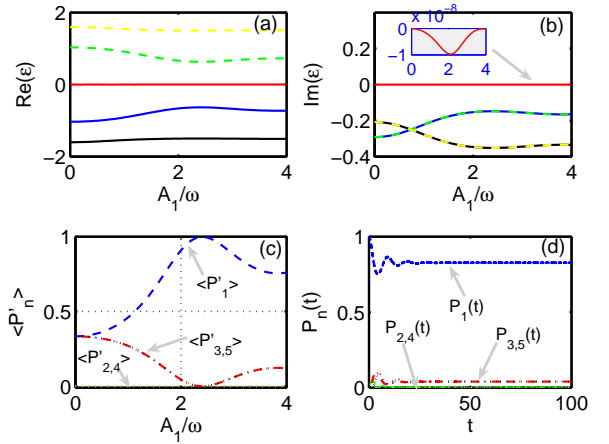}
\caption{(color online) 5-site system with $A_2=0$. (a-b) Real (left) and imaginary (right) parts of the quasienergies
as a function of $A_1/\omega$. Inset of panel (b) shows amplification  of imaginary  part of the zero-quasienergy (red line). (c) The time-averaged probability distribution
of the dark Floquet state corresponding to a nearly zero quasienergy in (a-b). (d) The detailed results for time evolution of all probabilities corresponding to the inset of panel (b) in Fig.~\ref{fig7}.
 Other parameters are the same as in the inset of panel (b) in Fig.~\ref{fig7}: $\omega=20,v=1$, and $(\alpha_2,\alpha_4)=(0,1)$.} \label{fig9}
\end{figure}

To further shed light on the physics of the system \eqref{c_n} with $N=5$ and $N=6$,
we present in Fig.~\ref{fig8}  the numerical equilibrium value $\langle P\rangle_{\rm{equ}}$ as functions of the driving parameters with integration time $t_f=100$.
Fig.~\ref{fig8} (a) represents the quantity $\langle P\rangle_{\rm{equ}}$ versus $A_1/\omega$ for the system with $N=5$, where the right-end site is undriven. We observe that for the parameter set $(\alpha_2,\alpha_4)=(0,1)$, the quantity $\langle P\rangle_{\rm{equ}}$ increases from $0.333$ as the driving parameter $A_1/\omega$ varies, and takes its maximum value of 1 at $A_1/\omega=2.4$.
Actually, the dependence of $\langle P\rangle_{\rm{equ}}$ on $A_1/\omega$  remains roughly unchanged even if the total evolution time is extended to six orders of magnitude (not listed here).
This provides a promising application that
the dissipative configuration with given parameter set $(\alpha_2,\alpha_4)=(0,1)$ may be used to keep its total population on a much higher level in the long-lived
steady state. We also note that for both parameter sets $(\alpha_2,\alpha_4)=(1,1)$ and $(\alpha_2,\alpha_4)=(1,0)$, the numerical equilibrium values $\langle P\rangle_{\rm{equ}}$ produced are the same and stay considerably smaller than that of $(\alpha_2,\alpha_4)=(0,1)$. Moreover, the quantities $\langle P\rangle_{\rm{equ}}$ produced for both $(\alpha_2,\alpha_4)=(1,1)$ and $(\alpha_2,\alpha_4)=(1,0)$ will become lower and lower if we extend the evolution time.
Fig.~\ref{fig8} (b) shows how the quantity $\langle P\rangle_{\rm{equ}}$ in a 6-site system varies as the
driving parameter $A_2/\omega$ is increased, assuming that the left-end site is undriven. We can observe in Fig.~\ref{fig8} (b) that, as seen previously in the
4-site system, the quantities  $\langle P\rangle_{\rm{equ}}$ for all sets of loss coefficient peak again at $A_2/\omega=2.4$-the zero of $J_0(A_2/\omega)$. In Fig.~\ref{fig9} (a-b), we plot the quasi-energies for the 5-site system with $(\alpha_2,\alpha_4)=(0,1)$ and with  other parameters as in Fig.~\ref{fig8} (a), and can note that, as like as the 3-site system, this 5-site system possesses a dark Floquet state with nearly zero
quasienergy and negligible population at all of the even $n$th
sites (see Fig.~\ref{fig9} (c)). A careful examination  reveals that the imaginary part of the seeming zero-quasienergy for $(\alpha_2,\alpha_4)=(0,1)$ is not a real zero but an extremely small negative nonzero number, which is of the order of $10^{-8}$ (see inset of Fig.~\ref{fig9} (b)), whereas for both $(\alpha_2,\alpha_4)=(1,1)$ and $(\alpha_2,\alpha_4)=(1,0)$, the Floquet-dark-state-related quasienergy's imaginary part is of order of $10^{-3}$ (see inset of Fig.~\ref{fig8} (a)), typically $10^{5}$ times larger than that of $(\alpha_2,\alpha_4)=(0,1)$. It is noted that the magnitude of the non-zero imaginary part of the quasienergy determines (inversely correlates to) the
temporal length of the system remaining in the corresponding  dark Floquet state without decay.  This  explains the reason why applying the periodic driving solely to the left-end site of  5-site system will give rise to
an equilibrium state with good stability
 over an enough longer evolution time  for the given set of loss coefficient $(\alpha_2,\alpha_4)=(0,1)$. In addition, we also plot in Fig.~\ref{fig9} (d) the time dependence of all probabilities for the corresponding 5-site system. As expected, the system has negligible population  at all of the even $n$th
sites during the dynamical evolution.

\begin{figure}[htp]
\center
\includegraphics[width=8cm]{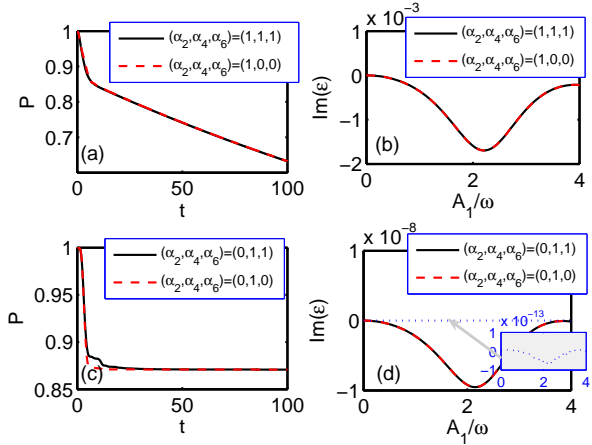}
\caption{(color online) 7-site system with $A_2=0$. (a) Time evolutions of the total probability for both parameter sets of loss coefficient $(\alpha_2,\alpha_4,\alpha_6)=(1,1,1)$ and  $(\alpha_2,\alpha_4,\alpha_6)=(1,0,0)$ with $A_1/\omega=2$, and starting the system with particle at site 1. (b) Imaginary  part of the nearly zero quasienergy for the associated dark Floquet state with the same parameters as in (a). (c) Time evolutions of the total probability for both $(\alpha_2,\alpha_4,\alpha_6)=(0,1,1)$ and  $(\alpha_2,\alpha_4,\alpha_6)=(0,1,0)$ with $A_1/\omega=2$, and starting the system with particle at site 1.
(d) Imaginary  part of the nearly zero quasienergy for the associated dark Floquet state with the same loss coefficients as in (c). In panel (b), we have incorporated the other case $(\alpha_2,\alpha_4,\alpha_6)=(0,0,1)$ as blue dotted lines in the main panel and inset (amplification). Other parameters are $\omega=20,v=1$.} \label{fig10}
\end{figure}

In what follows, we proceed to carry out the numerical simulation of model equation \eqref{c_n} with $N=7$, where the periodic driving is applied to address only the left-end site. As before, the particle is initially localized in the site 1. The results are shown in Fig.~\ref{fig10}, and we can expect to encounter
 the qualitatively similar results as in 5-site system. For both the parameter sets $(\alpha_2,\alpha_4,\alpha_6)=(1,1,1)$ and  $(\alpha_2,\alpha_4,\alpha_6)=(1,0,0)$ (in either case, the first lossy site from the left is site 2),
 the total populations show the same slow decay (see Fig.~\ref{fig10} (a)), both of which are closely related to
 the existence of dark Floquet states whose quasiernergies have small nonzero imaginary parts with the exactly same magnitudes (of order of $10^{-3}$) (see Fig.~\ref{fig10} (b)).
 By contrast, for both cases of $(\alpha_2,\alpha_4,\alpha_6)=(0,1,1)$ and  $(\alpha_2,\alpha_4,\alpha_6)=(0,1,0)$ (i.e., the first lossy site from the left is site 4), the dynamics are almost identical and
  the total populations are found to decay to the same steady values with long-term stability (see Fig.~\ref{fig10} (c)),
  which occurs as a consequence of the existence of dark Floquet states with exceedingly small nonzero imaginary parts of quasienergies (of order of $10^{-8}$, see the coincided curves of black  and red lines in Fig.~\ref{fig10} (d)).
  For comparison, we also consider the case of $(\alpha_2,\alpha_4,\alpha_6)=(0,0,1)$ (i.e., the case that the first lossy site from the left is site 6), and plot the imaginary part of quasienergy of the dark Floquet state as the blue-dotted lines in both the main panel and inset of Fig.~\ref{fig10} (d).
  We find that for $(\alpha_2,\alpha_4,\alpha_6)=(0,0,1)$, the associated dark Floquet state has a much smaller imaginary part (of order of $10^{-13}$), which means that the system will be driven to a steady state (i,e., the dark Floquet state) with much longer lifetime.

Taking both the 5-site and 7-site systems into consideration, our studies suggest that the numerical value of the system's effective decay depends not on how many lossy sites there are but on which of the lossy sites is nearest to the driven site.
When the first lossy site from the left is far enough away from the driven left-end site, the magnitude of the   imaginary part of quasienergy corresponding to dark Floquet state  will become extremely small, and the period of time for the system remaining in the stable dark Floquet state without decay will become exceedingly long. These notable features
are also supported by numerical studies of the systems with other odd numbers of
sites (not displayed here). Previous research has given a mathematical  proof that the dark Floquet state exists in all of the odd-$N$-state conservative systems, which is naturally validated for non-Hermitian
Hamiltonian systems\cite{Luo1}. Thus, for specially designed local dissipation, applying the periodic field only to the left-end site could drive the arbitrary odd-$N$-site  system to a long-lived steady state (i.e., stable dark Floquet state) which has a higher level of total probability (lower probability loss) as compared to the undriven case.

\section{conclusions}
We have discussed the effects of interplay between the periodic driving and localized dissipation for a
single quantum particle initially loaded in the left-end
site of
one-dimensional quantum lattice systems. To be concrete, we take the
number of lattice sites $N = 3, 5$, and 7 as  examples, and find that for an odd-$N$-site system,
there always exists a dark Floquet state whose quasienergy has an extremely small negative nonzero imaginary part, which will produce a nontrivial physical effect in dissipative systems. Such a feature is different from that of the conservative system where the quasienergy of the dark Floquet state is  not zero either but an extremely small nonzero all-real number, which, however, is negligible and has no observable physical effect. Furthermore,  we elucidate the prominent role played by the dark Floquet state in the suppression of decay in open quantum systems with localized dissipations. When subjected to localized dissipations from the even $n$th sites, the system will evolve in time toward the dark Floquet state, in which the population distributions among sites depend only on the driving parameters.
Specially, applying the periodic driving  only to left-end site  will produce a higher level of total population (lower probability loss) compared to the undriven case.
 The occurrence of nonzero imaginary part of the quasienergy means that the associated dark Floquet state will continue to decay with a very slow decaying rate.
We have discovered, fortunately, a nontrivial effect that the system's effective decay depends
not on the total number of lossy sites, instead only on the number of the first lossy lattice site from the driven left-end site. Particularly, the farther away from the driven left-end site the first lossy site from the left, the longer the time for the system staying in the dark Floquet state. The interplay between these effects enables the system to be driven to a long-lived  stable dark Floquet state with much higher level of total probability (lower probability loss), which opens additional possibilities for long-term suppression of decay in fully governable open quantum systems.

In addition, we have also numerically explored the  multi-site lattice systems with even numbers of sites. By studying the
number of lattice sites $N =4, 6$, we have found that by application of the periodic driving to the right-end site, the CDT effect makes it possible to switch off the tunneling between the  right-end site
and other undriven sites, and thus enable the system to be driven to a steady state without decay as in the undriven odd-number-site system. In essence, the existence of such a steady state  originated from the dark state, which emerges from the resulting effective undriven system with an odd number of sites, as the single driven right-end site is effectively removed from the dynamics due to the occurrence of CDT effect. It is noteworthy that the steady state, caused by the CDT between the driven right-end site and the remaining odd numbers of undriven sites, has a more stable overall probability over enough longer evolution time, in comparison with that caused by the CDT between the left-end site (initially
occupied by the particle) and its neighboring lossy site 2. This interesting finding indicates that
the CDT effect alone is not
enough to efficiently suppress the probability decay in the long-time dynamics, and the interplay (cooperation) between the effects of CDT and dark state is required.
We have considered only the simplest 4-site and 6-site system. In general, we should expect quantitatively similar results for other even numbers of lattice sites, controlled by local dissipation and periodic driving.

This investigation may be a first step toward understanding of the leading role played by the dark Floquet state in the suppression of decay in fully governable open quantum systems.
We have highlighted the essential differences of dark Floquet state in the dissipative systems with respect to
the conservative systems.
In view of the advancing experiments on the single-site addressability in
the optical lattices, where controlled  losses
can be made truly localized in selected sites of  optical
lattice\cite{Gericke,Bakr}, we expect that our results can be
tested in open quantum systems under currently available experimental conditions. Our conclusions are also applicable to a
large variety of systems. For instance, due to the equivalence between the Schr\"{o}dinger equation
and the optical wave equation, our results can be used to suppress the decay of optical signals in waveguides  with specially designed imaginary refractive index.

\acknowledgments
The work was supported by the National Natural Science
Foundation of China under Grants 11975110, 11764022,
11465009, the Scientific and Technological Research Fund of Jiangxi
Provincial Education Department (numbers GJJ180559, GJJ180581, GJJ180588),
and Open Research Fund Program of the State Key Laboratory of Low-
Dimensional Quantum Physics (KF201903).

Zhao-Yun Zeng and Lei Li
contribute equally to this work.

\end{document}